\documentclass[11pt]{article}
\usepackage{amssymb,slashed,graphicx,color,epsf,cite}
\usepackage{mathtools}
\usepackage{stackengine}

\stackMath
\usepackage{blkarray}
\usepackage{amsmath}
\usepackage{comment}
\usepackage{lscape}
\def\nn{\nonumber}

\def\be{\begin{equation}}
\def\ee{\end{equation}}
\def\ben{\begin{displaymath}}
\def\een{\end{displaymath}}
\def\bea{\begin{eqnarray}}
\def\eea{\end{eqnarray}}
\def\bm{\boldmath}

\makeatletter \@addtoreset{equation}{section} \makeatother

\begin{document}



\vspace{25pt}

\begin{center}

{\bf \Large {Ghost and Tachyon Free Propagation }\\
 \vspace{5pt} {\bf \Large {up to spin-3 in Lorentz Invariant Field Theories}}}

\vspace{0.3in}

{\large C. Marzo$^1$}

\vspace{0.3in}

\small{\textit{$^1${\it  Laboratory of High Energy and Computational Physics, NICPB, R\"{a}vala 10, 10143 Tallinn, Estonia}}}

\vspace{40pt}

\end{center}

\vskip 0.5in

\baselineskip 16pt

\begin{abstract}{We complete the set of spin-projector operators for fields up to rank-3 by providing all operators connecting sectors with same spin and parity. In this way we can broaden the search for unitary and non-tachyonic particle propagation in quadratic Lagrangian with inter-field mixing. We use the properties of projector algebra to reanalyze known theories and shed light towards new healthy ones. We do so with full control over the gauge constraints by working the form of the saturated propagator in an appropriate frame of reference.}
\end{abstract}

\vspace{15pt}

\thispagestyle{empty}

\pagebreak


\tableofcontents

\newpage

\section{Introduction}

The use of fields and local Lagrangians to fit the quantum behavior of relativistic particles has met unparalleled experimental support. In the effort to craft coherent quantum field models two main approaches can be followed which carry different views for the symmetries of the theory. In a up-bottom approach, symmetries are the starting point that shape the field interactions for a given particle content. In the lower spin sector, which can be described by low rank fields, the narrower space of invariants facilitates the building of such theories. Already for spin $\geq 2$, imposed symmetries lose their major constraining power, a paradigmatic case being the many different alternatives which accompany the simplest theory of gravity.
Aside of this mainstream attitude, a different scrutiny of quantum field models, more focused on particle's free propagation and consistency of their interactions, has been carried most notably in \cite{Gupta:1954zz,Deser:1969wk,Fang:1978rc,Barnich:1993vg,Barnich:1993pa,Barnich:1994mt,Barnich:1994db,Deser:1963zzc,Rivers:1964,VanNieuwenhuizen:1973fi}. This different focus has powerfully demonstrated how some symmetries can be brought into existence by the stronger claims of unitarity and causality of the quadratic Lagrangian. Moreover, in absence of obstructions, techniques have been developed to consistently extend the same symmetries allowing a non-linear completion, hence interactions. 
This bottom-up approach embraces the field formalism as a pathological carrier of multi-spin components, and selects constraints and symmetries over the generic Lorentz invariant Lagrangian to cure it. 
The path for a successful application of this method starts with, as said, an otherwise unconstrained Lorentz invariant quadratic action, built from a selected set of fields. A spectral analysis for this theory is then performed, which illuminates the presence of ghost-like pathological propagation. The nature of ghost being two-fold, both of Ostrogradsky type \cite{Ostro,Woodard:2015zca}\footnote{Notice that ways to welcome such ghosts have been explored \cite{Mannheim:2006rd,Anselmi:2018ibi,Anselmi:2018tmf,Donoghue:2019fcb}}, sourced by higher order derivative terms, or linked to the undefined nature of Lorentz metric. On top of this, tachyonic states also populates the propagator's poles and require care. After such polishing has been realized, which is in general a difficult task, a constrained quadratic action emerge. The art of building consistent interactions on top of linear system is highly aided by cohomological methods and by the less systematic, but more intuitive, Noether method \cite{Barnich:1993vg,Barnich:1993pa,Barnich:1994mt,Barnich:1994db,Hurth:1998nq}. This part will not be the subject of this paper which will focus on the constraining power of spectral analyses.   
To explore in a systematic way the particle spectrum of a theory it is favorable to explicit the link between fields, generally reducible representation of the Lorentz group, and particles, interpreted a' la Wigner as representations of the little group. Different methods have been developed, in particular in the context of gauge theories and higher-spin model building \cite{Fradkin:1977xi,Gao:2014fra,Burdik:2001hj,Metsaev:2013kaa,Buchbinder:2006nu,Moshin:2007jt,Buchbinder:2005ua,Fotopoulos:2008ka}, to pinpoint the physical degrees of freedom. We will rely on the techniques introduced by Rivers \cite{Rivers:1964}, and adopted in the seminal works \cite{VanNieuwenhuizen:1973fi,Neville:1978bk,Neville:1979rb,Sezgin:1979zf}, that exploit the algebra of projectors operators to highlight the spin-components of a given field. Even more relevantly this formalism solves smoothly the problem of the inversion of the equation of motion in presence of local symmetries, which are controlled by the gauge invariant \emph{saturated} propagator.     
These powerful techniques have handed the keys to reveal the inevitability of the Fierz-Pauli action \cite{VanNieuwenhuizen:1973fi} and reveal the particle spectrum of diffeomorphism invariant theories of gravity in first order formalism (for an incomplete list of works on the subject see \cite{Neville:1978bk,Neville:1979rb,Sezgin:1979zf,VanNieuwenhuizen:1973fi,Karananas:2014pxa,Sezgin:1981xs,Lin:2018awc,Percacci:2020ddy}). The latter case is paradigmatic of the use of many different fields with quadratic mixing and simultaneous transformation under gauge symmetry. It has received an increasing attention in the last years and has required the computation of projectors operators up to spin-3, an operation fully completed only recently \cite{Mendonca:2019gco,Percacci:2020ddy}.
When moving to higher-spin the simultaneous presence of many fields, possibly in an auxiliary role, is mandatory to provide the propagation of the wanted states. There is therefore fair expectation that future phenomenology could be tackled by models where particle propagation is encoded in multiple fields with collective gauge symmetries whose constraining power and renormalizability have not yet given the proper attention. This is particularly interesting given that consistent couplings of higher-spin particles seem to require higher derivative interactions and therefore a naive power-counting renormalization is no longer applicable. More promising, similarly to case of the non-linear sigma model, symmetries might help constraining the form of the UV terms and provide predictive models. \\
Surprisingly, while geometric theories of gravity have attracted most of the use of the method in question, almost none attention has been turned towards Lorentz invariant higher-spin propagation involving multiple fields, one notable exception being the study of Singh model in \cite{Mendonca:2019gco}.    
A possible reason to justify such lacking can be tracked to the missing operators needed to express the mixing terms involving all the fields. 
Up to field of rank-3, and therefore equal spin number, we provide the missing terms so to open the study of the spectrum for new unexplored models with possible collective symmetries. We illustrate a possible procedure, which has common roots with \cite{Karananas:2014pxa,Lin:2018awc}, with a greater focus around the explicit form of the saturated propagator and unambiguously establishes the nature of the propagating particles.\\
In this work we use the mostly minus metric signature. 
\section{Fields, particles and the quadratic action}

\subsection{Fields and particles}
While attempts to emancipate particle phenomenology from local fields are an intriguing line of past and present research \cite{Arkani-Hamed:2017jhn,Falkowski:2020fsu,Criado:2021itq,Criado:2021gcb}, their use is still mainstream when it comes to provide scattering amplitudes at relativistic energies. The capital issue of this approach is represented by the clash between the field components and the particle which they describe. More precisely, while tensor fields carry representations of the Lorentz group $\mathcal D\left(s_1,s_2\right)$\footnote{We limit ourselves to the bosonic case, avoiding double covering representations of the little group, suitable for fermions.}, the particle's labels are linked to the little group. These are the smaller $SU(2)$ for massive particles, and $U(1)$ for massless, in terms of which $\mathcal D\left(s_1,s_2\right)$ is reducible. In the massive case the physical relevant quantum number is therefore the spin $s$, with representations $2s + 1$ dimensional and values $-s, -s+1, ... s-2, s-1, +s$. Similarly, for the little group of massless particles, the representations are also identified with the integer number $s$, connected to the eigenvalues of the conserved helicity operator, with only two different states $\pm s$. 
When parity invariance is meaningful, also the twofold discrete values of the parity operator ($P = \pm 1$) aid in giving a full description of Poincar\'e invariant systems through fields, with the abelian subgroup of translations being trivially realized by their continuous momentum dependence.
In our analysis, as customary in the literature, we will use the decomposition of fields in the $SO(2)$ little group which is not only enough to derive conclusions about the causal and unitary propagation of massive particles, but also provides the intermediate step in drawing the unitarity of helicity states within the formalism of the saturated propagator. It is a by now standard notation to label the little group components within the reducible Lorentz tensor with the symbol $S^p$, with $S$ the $SU(2)$ spin number and $p$ the parity eigenvalue. To this symbol we add a further subscript to distinguish multiple components of the same $S^p$ element for the totality of cases analyzed in this paper, from rank-0 to rank-3 fields. 
Therefore, following the enumeration of \cite{Percacci:2020ddy} and extending it to include the vector and scalar case, we have: 
\begin{eqnarray}\label{eq1}
&\phi_{\mu \nu \rho} \supset & 3_1^- \oplus 2_1^+ \oplus 2_2^+ \oplus 2_3^+ \oplus 2_1^- \oplus 2_2^- \oplus 1^+_1 \oplus 1^+_2 \oplus 1^+_3 \oplus 1^-_1 \oplus 1^-_2 \oplus 1^-_3 \oplus  \nn \\ && \oplus 1^-_4 \oplus 1^-_5 \oplus 1^-_6 \oplus  0^+_1 \oplus 0^+_2 \oplus 0^+_3 \oplus 0^+_4 \oplus 0^-_1 , \nn \\
&\phi_{\mu \nu} \supset & 2_4^+ \oplus 1^-_7 \oplus 0^+_5 \oplus 0^+_6 , \nn \\
&\phi_{\mu} \supset & 1^-_8 \oplus 0^+_7 ,\nn \\
&\phi_{} \supset & 0^+_8 
\end{eqnarray}
where we only impose symmetry for the rank-2 tensor.
Eq~(\ref{eq1}) makes explicit how a single indexed field carries multiple particles, with substantial growth in their number with the field rank. Therefore, when using fields as building blocks for the dynamics of higher-spin particles, the necessity of constraining the fields arises. Fierz and Pauli \cite{Fierz:1939ix} presented the rules to build linear equations of motion which only propagate a spin $s$ particle adopting a rank-$s$ symmetric and traceless tensor field $\Phi_{\mu_1 \mu_2 \cdots \mu_s}$, which therefore carries the Lorentz representation $\mathcal{D}(s/2,s/2)$ .
Being such single representation already reducible in $SU(2)$ components, on top of the Klein-Gordon equation
\begin{eqnarray}
\left(\Box + m^2\right)\phi_{\mu_1 \mu_2 \cdots \mu_s} = 0 ,
\end{eqnarray}
also the null-divergence condition 
\begin{eqnarray}
\partial^{\mu_i}\phi_{\mu_1 \mu_2 \cdots \mu_s} = 0  \,\,\,\,\,\, \left(i = 1,..s\right) ,
\end{eqnarray}
must be imposed to prevent the propagation of particles with spin lower than $s$. 
It was immediately realized by Fierz and Pauli that the simplicity of their universal description for all integer spin particle could not survive the introduction of interactions without leading to inconsistencies. Such inconsistencies could be healed by relaying to a Lagrangian origin of the equations, enriched with auxiliary, lower rank, fields. 
Since the seminal Fierz-Pauli paper, most of the subsequent attempts to find healthy Lagrangians for propagation of higher-spin particles relied on quadratic mixing among different fields. For instance, taking the simple but already involved case of a spin-3 particle, the Singh-Hagen \cite{Singh:1974qz} proposal uses a Lagrangian with a collective rank-3 and rank-0 dynamics, while the Klishevich-Zinoviev model \cite{Klishevich:1997pd,Zinoviev:2008ck} needs all fields with rank less or equal than 3. 
We will study, as exemplary cases, such models with our complete set of projectors in the following pages \footnote{The unitarity of the Singh-Hagen model already investigated in \cite{Mendonca:2019gco} where, to only use rank-3 projectors, a clever integration of the auxiliary rank-0 field was employed, trading the linear Lagrangian to a non-linear one.}.  

\subsection{Projector operators}
Having recognized the relevance of Lagrangian theories with multiple fields entangled in quadratic mixing, we recap now the efficient approach based on projector operators\cite{Mendonca:2019gco,Neville:1978bk,Neville:1979rb,Sezgin:1979zf,VanNieuwenhuizen:1973fi,Karananas:2014pxa,Sezgin:1981xs,Lin:2018awc,Percacci:2020ddy} to assess the nature of their particle spectrum.
In what follows we will, when stating general properties of the action, hide the index structure and consider a index-less superfield formalism as in $\Phi = \left\lbrace \phi_{\mu \nu \rho}, \phi_{\mu \nu}, \phi_{\mu}, \phi  \right\rbrace$, considering field contractions in the intuitive form 
\bea
\Phi \, \Psi = \phi_{\mu_1 \mu_2 \mu_3}\,\psi^{\mu_1 \mu_2 \mu_3}+ \phi_{\mu_1 \mu_2} \,\psi^{\mu_1 \mu_2} + \phi_{\mu_1}\,\psi^{\mu_1} + \phi\, \psi \, ,
\eea
and similarly for supermatrices $K$ 
\bea
&&\Phi \, K \, \Psi = \nn \\ &&\phi_{\mu_1 \mu_2 \mu_3} \,\kappa^{\mu_1 \mu_2 \mu_3 \,\,\nu_1 \nu_2 \nu_3}\, \psi_{\nu_1 \nu_2 \nu_3} + \phi_{\mu_1 \mu_2 \mu_3} \,\kappa^{\mu_1 \mu_2 \mu_3 \,\,\nu_1 \nu_2}\, \psi_{\nu_1 \nu_2} + \phi_{\mu_1 \mu_2} \,\kappa^{\mu_1 \mu_2 \,\,\nu_1 \nu_2 \nu_3}\, \psi_{\nu_1 \nu_2 \nu_3} + \cdots \nn \\
&&\quad \quad \cdots \phi_{\mu_1} \,\kappa^{\mu_1 \,\,\nu_1}\, \psi_{\nu_1} + \phi_{\mu_1} \,\kappa^{\mu_1}\, \psi + \phi_{} \,\kappa^{\nu_1}\, \psi_{\nu_1} + \phi \, \kappa \, \psi \,\,, 
\eea
where, of course, in practical examples some components might be absent. 

The starting point of the algorithm is the source-dependent quadratic action, which in momentum space can be put in the economical form 
\begin{equation}\label{eq2}
\mathcal S =  \frac{1}{2} \int d^4 q \, \bigg( \Phi(-q) \,  \mathcal \, K(q) \, \Phi(q) + \mathcal{J}(-q)\Phi(q) + \mathcal{J}(q)\Phi(-q) \bigg)\, . 
\end{equation}
The explicit particle content of (\ref{eq2}) is accessed by expanding the fields in terms of the irreducible $S^p$ components, a process taken care by a set of operators with quite an involved superscript and subscript index structure:
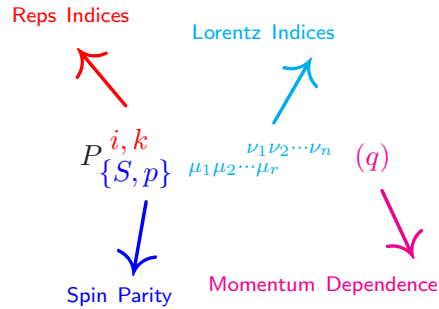
\[
  \def\stackalignment{r}
  \displaystyle{P}^{\,\, {\color{red}\stackon{%
    \stackon{\displaystyle  i,k}{%
      \scalebox{3}{\rotatebox{+40}{$\uparrow$}}\,}%
    }{\scriptstyle\mathsf{Reps~Indices}~~}}}%
    _{{\color{blue} \stackunder{%
    \stackunder{\displaystyle \left\lbrace S,p\right\rbrace}{%
      \scalebox{3}{\rotatebox{-10}{$\downarrow$}}\,}%
    }{\scriptstyle\mathsf{Spin~Parity}}}}%
    \,\,\,
    {\color{cyan} \stackon{%
    \stackon{%
     \displaystyle{{}_{\mu_1 \mu_2 \cdots \mu_r}^{\quad\quad \nu_1 \nu_2 \cdots \nu_n}}}{%
      \scalebox{3}{\rotatebox{-30}{$\uparrow$}}\,}%
      }{\scriptstyle{\mathsf{Lorentz~Indices}}}}%
           ~~{\color{magenta} \stackunder{%
    \stackunder{%
     \displaystyle{\left(q\right)}~~~~~}{%
      \scalebox{3}{\rotatebox{+25}{$\downarrow$}}\,}%
      }{\scriptstyle{\mathsf{Momentum~Dependence}}}}%
\]\bea
\label{eq3}
\eea
While is difficult to top the clear illustrations given by \cite{Neville:1978bk,VanNieuwenhuizen:1973fi,Karananas:2014pxa,Sezgin:1981xs,Lin:2018awc,Percacci:2020ddy} we briefly review the main properties of the operators in (\ref{eq3}) to keep these pages as self-contained as possible.
We start by the simple case of a single field $\phi_{\mu_1 \mu_2 \cdots \mu_n}\left(q\right)$, then the subset of operators with $i=k$ in ~(\ref{eq3}) are actually projectors onto the space of spin $S$ and parity $p$, fulfilling the completeness relation
\bea \label{eq4}
\phi_{\mu_1 \mu_2 \cdots \mu_n}\left(q\right) = \sum_{S,p,i} P^{i,i}_{\left\lbrace S,p\right\rbrace}{}_{\mu_1 \mu_2 \cdots \mu_n}^{\quad\quad \nu_1 \nu_2 \cdots \nu_n}\left(q\right) \phi_{\nu_1 \nu_2 \cdots \nu_n}\left(q\right).
\eea
The meaning of the \emph{Reps Indices} $i,i$ is therefore to keep track of the multiple representations within the spin/parity sector $\left\lbrace S,p\right\rbrace$, as enumerated in eq.~(\ref{eq1}). The need of a  double entry is then intended when considering \emph{transitions} between different representation with same $S$ and $p$ values. Being possible to realize such transition also among fields of different rank, we have considered in~(\ref{eq3}) the generic case with different number of upper and lower Lorentz indices. For instance, by looking at eq.~(\ref{eq1}), we might recover a vector with spin/parity $1^-$ out of a rank-3 tensor via   
\bea \label{eq5}
 P^{8,1}_{\left\lbrace 1,-\right\rbrace}{}_{\mu}^{\,\,\,\nu_1 \nu_2 \nu_3}\,\, \phi_{\nu_1 \nu_2 \nu_3} =  P^{8,1}_{\left\lbrace 1,-\right\rbrace}{}_{\mu}^{\,\rho_1 \rho_2 \rho_3} P^{1,1}_{\left\lbrace 1,-\right\rbrace}{}_{\rho_1 \rho_2 \rho_3}^{\quad\nu_1 \nu_2 \nu_3} \,\,\phi_{\nu_1 \nu_2 \nu_3},
\eea
where we neglected the momentum dependence to lighten the notation. 
Equations~(\ref{eq4}-\ref{eq5}) are particular applications of
\bea \label{eq6}
&&\sum_{S,p,i} P^{i,i}_{\left\lbrace S,p\right\rbrace}{}_{\mu_1 \mu_2 \cdots \mu_n}^{\quad\quad \nu_1 \nu_2 \cdots \nu_n} = \displaystyle{\hat 1}{}_{\mu_1 \mu_2 \cdots \mu_n}^{\quad\quad \nu_1 \nu_2 \cdots \nu_n} , \nn \\ &&\nn \\
&& P^{i,k}_{\left\lbrace S,p\right\rbrace}{}_{\mu_1 \mu_2 \cdots \mu_n}^{\quad\quad \rho_1 \rho_2 \cdots \rho_n} \,\, P^{j,w}_{\left\lbrace R,m\right\rbrace}{}_{\rho_1 \rho_2 \cdots \rho_n}^{\quad\quad \nu_1 \nu_2 \cdots \nu_n} = \delta_{k, j}\, \delta_{S, R}\, \delta_{p, m}\, P^{i,w}_{\left\lbrace S,p\right\rbrace}{}_{\mu_1 \mu_2 \cdots \mu_n}^{\quad\quad \nu_1 \nu_2 \cdots \nu_n} , \nn \\ &&\nn \\
&& P^{i,j}_{\left\lbrace S,p\right\rbrace}{}_{\mu_1 \mu_2 \cdots \mu_n}^{\quad\quad \nu_1 \nu_2 \cdots \nu_n} = 
\left(P^{j,i}_{\left\lbrace S,p\right\rbrace}{}^{\nu_1 \nu_2 \cdots \nu_n}_{\quad \quad \mu_1 \mu_2 \cdots \mu_n}\right)^*  ,
\eea
which describe completeness of the projectors, orthogonality and hermitianicity of the operators \footnote{We introduced the unit operator $\hat{1}$ in the field space of given rank and index symmetry so that, for instance, for a symmetric rank-2 fields we have $\hat{1}_{\mu_1 \mu_2}^{\quad \, \nu_1 \nu_2} = \frac{1}{2}\left(\delta^{\nu_1}_{\mu_1}\delta^{\nu_2}_{\mu_2}+\delta^{\nu_2}_{\mu_1}\delta^{\nu_1}_{\mu_2}\right)$.}.
The task of computing the explicit form of these operators, for the set of rank-2 and rank-3 fields has been only recently completed \cite{Mendonca:2019gco,Percacci:2020ddy}. This paper's original contribution is to provide the missing operators to study the mixing between all the fields up to rank-3, including the scalar-tensor and vector-tensor transition operators. 
We refer to the cited literature for details about the computation of the actual projectors, in particular for the role of the transversal and longitudinal operators which allows possible future generalizations towards higher spins. For our computation a brute force method has been used, with an appropriate code to generate tensor covariant combinations with given symmetries so to impose eq.~(\ref{eq6}) and solve the corresponding linear system. We provide a shortened list of such operators, which can be completed to the full set with the use of (\ref{eq6}), in the Appendix. The full code-ready set can instead be found in the ancillary file. 
For this purpose the following tools have been of great help \cite{xact:20,Brizuela:2008ra,Frob:2020gdh}. 
Once the set of operators fulfilling eq.~(\ref{eq6}) is known, the investigation about the nature of the particle spectrum follows almost mechanically. In particular, within this formalism, the subtle presence of gauge symmetries can be revealed and dealt with in a simple and physically intuitive way. 
\subsection{The quadratic action and the saturated propagator}
The source-dependent quadratic action eq.~(\ref{eq2}) can be now manipulated to reveal the propagating spin/parity sectors via the expansion
\begin{eqnarray}\label{eq8}
&&\int d^4 q \,  \Phi(-q) \, K(q) \, \Phi(q) \, =  \int d^4 q \, \Phi(-q) \, \sum_{S,p,i,j} \bigg( a^{\left\lbrace S,p \right\rbrace}_{i,j}  P^{i,j}_{\left\lbrace S,p \right\rbrace} \bigg) \, \Phi(q)  \,\,. 
\end{eqnarray}
To get the fundamental matrices $a^{\left\lbrace S,p \right\rbrace}_{i,j}$ we need to get the component of the supermatrix $K(q)$ corresponding to the fields which include the $i$-nth and $j$-nth representation of spin/parity $S^p$ (\ref{eq1}).
The Lorentz indices of this object are then traced against the operator $ P^{i,j}_{\left\lbrace S,p \right\rbrace}$, opportunely weighted with the dimension of the $S^p$ representation $d_S = 2\,S+1$. So, for instance, to get $a^{\left\lbrace 1,- \right\rbrace}_{1,8}$ we do 
\bea
a_{1,8}^{\left\lbrace 1,- \right\rbrace} = \frac{1}{3} \,\, P^{1,8}_{\left\lbrace 1,- \right\rbrace}{}_{\mu_1 \mu_2 \mu_3}^{\quad \quad  \,\, \nu_1} \,\,\kappa^{\mu_1 \mu_2 \mu_3}_{\quad \quad  \,\, \nu_1} \,,
\eea
or similarly for diagonal elements
\bea
a_{4,4}^{\left\lbrace 2,+ \right\rbrace} = \frac{1}{5} \,\, P^{4,4}_{\left\lbrace 2,+ \right\rbrace}{}_{\mu_1 \mu_2}^{\quad  \,\, \nu_1 \nu_2} \,\, \kappa^{\mu_1 \mu_2}_{\quad \,\, \nu_1 \nu_2} \,,
\eea
and so on. The matrices $a^{\left\lbrace S,p \right\rbrace}_{i,j}$ are central in these spectral investigations, encoding all the information of the particle quanta in a form easy to manipulate. Indeed, with the form (\ref{eq8}) the computation of the propagator $\mathcal D(q)$ in presence of the sources $J(q)$ translates in the inversion of the equation 
\bea \label{eq10}
&& K(q) \, \mathcal D(q) \, =   \sum_{S,p,i,j} \bigg( a^{\left\lbrace S,p \right\rbrace}_{i,j}  P^{i,j}_{\left\lbrace S,p \right\rbrace} \bigg) \, \mathcal D(q) =  \hat{1} \,\,, 
\eea
which has the solution
\bea \label{eq10b}
\mathcal D(q) = \sum_{S,p,i,j}  b^{\left\lbrace S,p \right\rbrace}_{i,j}  P^{i,j}_{\left\lbrace S,p \right\rbrace} \,\,,
\eea 
with $b^{\left\lbrace S,p \right\rbrace}_{i,j} = \left(a^{\left\lbrace S,p \right\rbrace}_{i,j}\right)^{-1}$. The advantage in using such decomposition is the direct connection with gauge symmetries, which arise when the determinant of $a^{\left\lbrace S,p \right\rbrace}_{i,j}$ is zero, making the naive inversion impossible. 
Then, in presence of a set of $n$ null-vectors $X_i^{s = 1,2\cdots n}$ for the matrix $a^{\left\lbrace S,p \right\rbrace}_{i,j}$, we have corresponding $n$ symmetries under the gauge transformations 
\bea \label{eq11}
\delta \Phi = X^s_i \, P^{i,j}_{\left\lbrace S,p \right\rbrace} \Psi \,, \quad\,\,(s = 1,2\cdots n)\,,
\eea
with arbitrary superfield $\Psi$. To fix such gauge freedom we take the non-degenerate submatrix $\tilde{a}^{\left\lbrace S,p \right\rbrace}_{i,j}$, which can be promptly inverted. Then \emph{we can recover gauge invariance} asking the sources to solve for the gauge constraints 
\bea \label{eq12}
X^{*}{}^s_j \,\, P^{i,j}_{\left\lbrace S,p \right\rbrace} \,\, J \left(q\right) = 0 \,, \quad\,\,(s = 1,2\cdots n)\,,
\eea
and defining   
\bea \label{eq13}
\mathcal D_S(q) =  \tilde{J}^*\left(q\right) \bigg(\sum_{S,p,i,j}  \tilde{b}^{\left\lbrace S,p \right\rbrace}_{i,j}  P^{i,j}_{\left\lbrace S,p \right\rbrace}\bigg)  \tilde{J}\left(q\right) 
\eea 
where we used $\tilde J \left(q\right)$ to refer to solutions of eq.~(\ref{eq12}). 
\subsection{Poles, ghost and tachyons}
Definition (\ref{eq13}) is the gauge invariant \emph{saturated propagator} which has the generic structure of an undefined quadratic form in the sources with possible poles in the squared momentum $q^2$. In proximity of such poles we recover the structure 
\bea \label{eq14}
\mathcal  \displaystyle{\lim_{q^2 \to M_a^2}} D_{S}\left(q\right) \sim \frac{\sum_{m} r_m | j^m|^2}{\left(q^{2}-M_a^{2}\right)^n} \,\,, 
\eea 
where $j^m$ are linear combinations of source components with definite properties under angular momentum ($M_a\neq 0$) or helicity ($M_a = 0$) transformations (for details see section II of \cite{Schwinger:1970xc}). Many factors can concur to challenge the healthy propagation of particles, providing in turn strong constraints over the parameter space of the linear model. As known, the residue of the propagator poles has a special role in defining the unitarity of the model and the undefined Lorentz metric cannot assure the positivity of the residue in (\ref{eq14}), thus generally propagating \emph{ghosts}. Connected to this is also the elimination of possible poles of order greater than 1 in $q^2$, linked to Ostrogradsky instabilities ($n>1$ in eq.~(\ref{eq14})).
The tribute to causality is instead paid by avoiding superluminal propagation which is triggered by complex masses $M^2_a < 0$ providing yet another constraint to account for. \\
A certain number of shortcuts have been developed to avoid dealing with (\ref{eq14}) directly and instead reduce the problem to the simpler matrices $\tilde{b}^{\left\lbrace S,p \right\rbrace}_{i,j}$.
Indeed, by restricting ourselves to non-degenerate massive particles, the positivity of the quadratic form over the pole $q^2 \rightarrow M^2$ can be read off the simple formula
\label{eq14b}
\[
\stackunder{Res}{q^2 \rightarrow M^2}\quad\,\sum_i\, \left(-1\right)^p\,\tilde{b}^{\left\lbrace S,p \right\rbrace}_{i,i} \,> 0 \,.
\]
This approach misses potentially interesting scenarios as massless propagation, critical cases where different spin sectors share a common mass and, notably for our work , collective gauge invariant descriptions of massive particles (see section \ref{KZ}). For this reason, in the spirit of \cite{VanNieuwenhuizen:1973fi,Lin:2018awc} we will directly rely on the source-dependent saturated propagator to derive any conclusion about unitarity and causality. The impairment to get such information from  eq.~(\ref{eq14}) is connected to the imposition of constraints to the sources, which will be difficult to exhibit by keeping an index-free formalism. Therefore some progress can be made by exploiting Lorentz invariance and solve the gauge constraints in particular frames. In the massive case we feed our algorithm with the form of (\ref{eq12}) in the frame $q = (\omega,\vec{0})$ and solve for the components, then such subset of independent sources is reinserted in (\ref{eq12}) where the residue is easily computed. The massless case is slightly more subtle because of spurious poles in $q^2$ are also present in the polarization operators. Therefore we intermediately use the frame $q = (\omega,0,0,\kappa)$ and only later impose the light-like limit $\kappa \rightarrow \omega$. 
\section{Applications: review}
%
Before adopting the projectors technology to explore uncharted territories, we review more familiar ones stressing, with the benefits of insight, the role of quadratic mixing and gauge symmetries. 
\subsection{Proca-Stueckelberg-Goldstone} \label{secPSG}
The interacting theory of the vector and scalar field is a simple yet rich playground which shows the interplay between our formalism and the physical properties of particles they accommodate. In this section, we will extensively revisit Proca and Stueckelberg theories. This subject has been already exposed in \cite{Lin:2019ugq}, but we find it necessary to present it here as well, before moving to more complex scenarios.
We start analyzing the massive vector system described by the Proca action 
\bea
\mathcal S_{P} = \int d^4 x\left[-\frac{1}{4}\left(\partial_{\mu}V_{\nu}-\partial_{\nu}V_{\mu}\right)\left(\partial^{\mu}V^{\nu}-\partial^{\nu}V^{\mu}\right) + \frac{m_V^2}{2}V_{\mu}V^{\mu}\right] \,.
\eea
The decomposition (\ref{eq1}) defines two spin/parity sectors $1^-$ and $0^+$ with the (one-dimensional) matrices
\bea
&& a^{\left\lbrace 1,- \right\rbrace}_{8,8} =  q^2 - m_V^2 \,,\nn \\
&& a^{\left\lbrace 0,+ \right\rbrace}_{7,7} =  m_V^2 .
\eea
The mass parameter removes the possible degeneracy of the scalar sector, thus breaking the corresponding gauge symmetry. The saturated propagator $\mathcal D_S(q) =  \mathcal D^{1^-}_S(q) + \mathcal D^{0^+}_S(q)$ can therefore be trivially computed as 
\bea
&&\mathcal D^{1^-}_S(q) = J^{*}_{\mu}(q)\,\,\tilde{b}^{\left\lbrace 1,- \right\rbrace}_{8,8} \,\, P_{\left\lbrace 1^-\right\rbrace}^{8,8}{}^{\mu \nu} \,\,J_{\nu}(q) = \frac{J^*{}^{\mu}(q)J^{\nu}(q)}{q^2-m_V^2}\left(\frac{q_{\mu}q_{\nu}}{q^2}-g_{\mu \nu}\right)\,\, ,\nn \\
&&\mathcal D^{0^+}_S(q) = J^{*}_{\mu}(q)\,\,\tilde{b}^{\left\lbrace 0,+ \right\rbrace}_{7,7} \,\, P_{\left\lbrace 0^+\right\rbrace}^{7,7}{}^{\mu \nu} \,\,J_{\nu}(q) = \frac{J^*{}^{\mu}(q)J^{\nu}(q)\,q_{\mu}q_{\nu}}{m_V^2 q^2} \,\,.
\eea  
We see that, while the $0^+$ sector has only a spurious massless pole, the sector $1^-$ might show a healthy, non-tachyonic, propagation. Going to the frame $q^{\mu} = (\omega, \vec{0})$ we simply find
\bea \label{eq15}
\mathcal  \displaystyle{\lim_{\omega^2 \to m_V^2}} D_{S}\left(q\right) = \frac{|J^{1}|^2+|J^{2}|^2+|J^{3}|^2}{\omega^{2}-m_V^{2}} \,\,, 
\eea 
with the three degrees of freedom as expected for a massive spin-1 particle. 
It is known that, to circumvent the bad $q^2 \rightarrow \infty$ behavior of the $0^+$ sector, gauge symmetries must be introduced to sweep it away. In \emph{all known cases} this will require an extension with quadratic mixing between the vector and a scalar sector. In the minimal case this can be achieved via Stueckelberg mechanism, with only one field introduced in the simple 
\bea
\mathcal S_{S} = \mathcal S_{P} + \int d^4 x \,\left[\frac{1}{2} \,\, \partial_{\mu}\phi_S \, \partial^{\mu}\phi_S + m_V \,\phi_S \,\partial_{\mu} V^{\mu} \right]  \,.
\eea
We see the appearance of quadratic mixing in order to entangle the two fields in a common symmetry. This symmetry can be revealed noticing that, now, the sector $0^+$ is enhanced to the \emph{degenerate} $2\times2$ matrix
\bea \label{eq15b}
a^{\left\lbrace 0,+ \right\rbrace}_{i,j} = 
\begin{pmatrix}
a_{7,7} & a_{7,8} \\
a_{8,7} & a_{8,8} 
\end{pmatrix}
=
\begin{pmatrix}
m_V^2 & i\,m_V \sqrt{q^2}  \\
- i\, m_V \sqrt{q^2} & q^2 
\end{pmatrix} \, .
\eea
A direct application of (\ref{eq11}) then reveals, in momentum space, the symmetry 
\bea \label{eq16}
\delta \Phi = \left(\delta V_{\mu} , \delta \phi_S \right) =  \left(i\, q_{\mu}\,\psi, m_V \psi \right),
\eea
with $\psi$ an arbitrary scalar field.
We can fix the freedom of (\ref{eq16}), to allow the computation of the propagator, by arbitrary choosing  either diagonal element in (\ref{eq15b}) as non-degenerate submatrix. The equivalence of both choices is granted by the source constraint (\ref{eq12}) which becomes 
\bea
i\,\frac{q_{\mu}}{m_V}\,J^{\mu}(q) = J(q) \,,
\eea
where the indexless $J(q)$ is the source for the scalar field $\phi$ \footnote{We hope to curb the inevitable profusion of symbols by leaving the number of indices to signal to which field a given source is connected to.}. Besides the role of the gauge invariance, the saturated propagator, in the limit $q^2 = \omega^2 \rightarrow m_V^2$, is formally identical to the Proca one, propagating three states of the massive $1^-$ spin sector. 
We stress that such spectral analysis displays features also included by more involved Lagrangians with a Stueckelberg-like quadratic part. In particular, the presence of quadratic mixing between fields of different rank, and the corresponding inhomogeneous transformations required by gauge invariance, are manifested by Goldstone states in theories with spontaneously symmetry breaking. 
\subsection{Einstein-Palatini} \label{secEP}
The massless Fierz-Pauli Lagrangian for a symmetric rank-2 field propagates 2 helicity states as shown, with the help of the projectors algebra, in \cite{VanNieuwenhuizen:1973fi}. There, the apparent propagation of a ghost-like spin-0 particle is shown to disappear in the limit $q^2 \rightarrow 0$ owing to the constrained sources. As an interesting application we revisit the same problem adopting a rank-3 tensor as an auxiliary field in the so called Palatini formulation. To the best of our knowledge the saturated propagator for this formulation has been explicitly worked out, in coordinate space, only in \cite{Karananas:2014pxa}. On top of the many interesting applications for the first-order formulation of the gravitational problem, it also gives a non-trivial display of a model requiring a rank-3 and rank-2 field cooperation in order to propagate a single helicity-2 particle. \\
The quadratic action we target is of the form  
\bea
\mathcal S_{EP} &=& a_{EP}\, \int d^4 x \left(\delta^{\mu \nu} + H^{\mu \nu} \right)\left(\partial_{\alpha}A^{\alpha}_{\,\,\mu \nu} - \partial_{\nu}A^{\alpha}_{\,\,\mu \alpha} + A^{\alpha}_{\,\,\nu \mu}A^{\beta}_{\,\,\alpha \beta} - A^{\alpha}_{\,\,\beta \mu}A^{\beta}_{\,\,\alpha \mu} \right) \nn \\
&=&  a_{EP}\, \int d^4 x \left[H^{\mu \nu}\left(\partial_{\alpha}A^{\alpha}_{\,\,\mu \nu} - \partial_{\nu}A^{\alpha}_{\,\,\mu \alpha}\right) + A^{\alpha \,\,\,\,\mu}_{\,\,\mu}A^{\beta}_{\,\,\alpha \beta} - A^{\alpha}_{\,\,\beta \mu}A^{\beta\,\, \mu}_{\,\,\alpha}  + O\left(H \, A^2\right)\right] \,, \nn \\ 
\eea
and we require $H^{\mu \nu} = H^{\nu \nu}$  and $A^{\alpha}_{\,\,\mu \nu} = A^{\alpha}_{\,\,\nu \mu}$, the latter equality to avoid many irrelevant spin representations. For this set of fields eq. (\ref{eq1}) is now reduced to  
\begin{eqnarray}\label{eq18}
&A_{\mu \nu \rho} \supset & 3_1^- \oplus 2_1^+ \oplus 2_2^+ \oplus 2_1^- \oplus 1^+_1 \oplus 1^-_1 \oplus 1^-_2 \oplus 1^-_4 \oplus 1^-_5  \oplus  0^+_1 \oplus 0^+_2 \oplus 0^+_4 , \nn \\
&H_{\mu \nu} \supset & 2_4^+ \oplus 1^-_7 \oplus 0^+_5 \oplus 0^+_6 \,.
\end{eqnarray}
In terms of these representations the relevant spin/parity matrices become
\bea \label{eq18a}
a^{\left\lbrace 2,+ \right\rbrace}_{i,j} &=& 
a_{EP}
\begin{pmatrix}
-2 & 0 & -i \frac{\sqrt{q^2}}{3}  \\[6pt]
0 & 1 & -i \sqrt{\frac{2}{3}}\sqrt{q^2}  \\[6pt]
i \frac{\sqrt{q^2}}{3} & i \sqrt{\frac{2}{3}}\sqrt{q^2} & 0 
\end{pmatrix} \, , \\ \nn \\
a^{\left\lbrace 1,- \right\rbrace}_{i,j} &=& 
a_{EP}
\begin{pmatrix}
\frac{4}{3} & -\frac{\sqrt{5}}{3} & \frac{2\sqrt{5}}{3} & -\frac{1}{3}\sqrt{\frac{5}{2}} & i \sqrt{\frac{5 q^2}{6}} \\[6pt]
-\frac{\sqrt{5}}{3} & -\frac{1}{3} & -\frac{1}{3} & -\frac{2 \sqrt{2}}{3} & i \sqrt{\frac{q^2}{6}} \\[6pt]
\frac{2\sqrt{5}}{3} & -\frac{1}{3} & -\frac{4}{3} & -\frac{1}{3\sqrt{2}} & -i \sqrt{\frac{q^2}{6}} \\[6pt]
-\frac{1}{3}\frac{\sqrt{5}}{2} & -\frac{2\sqrt{2}}{3} & -\frac{1}{3\sqrt{2}} & \frac{1}{3} & -i \sqrt{\frac{q^2}{2\sqrt{3}}} \\[6pt]
- i \sqrt{\frac{5 q^2}{6}} & -i \sqrt{\frac{q^2}{6}} & i \sqrt{\frac{q^2}{6}} &  i \sqrt{\frac{q^2}{2\sqrt{3}}} & 0
\end{pmatrix} \, ,
\eea
and 
\bea \label{eq18c}
a^{\left\lbrace 0,+ \right\rbrace}_{i,j} = 
a_{EP}
\begin{pmatrix}
0 & \frac{1}{\sqrt{2}} & 2 & - i \frac{\sqrt{q^2}}{\sqrt{3}} & i \sqrt{q^2} \\[6pt]
\frac{1}{\sqrt{2}} & -1 & \frac{1}{\sqrt{2}} & - i \sqrt{\frac{2 q^2}{3}}  & - i \sqrt{\frac{q^2}{2}} \\[6pt]
2 & \frac{1}{\sqrt{2}} & 0 & 0 & 0 \\[6pt]
i \sqrt{\frac{q^2}{3}} & i \sqrt{\frac{2 q^2}{3}} & 0 & 0 & 0 \\[6pt]
- i \sqrt{q^2} & i \sqrt{\frac{q^2}{2}} & 0 & 0 & 0
\end{pmatrix} \, .
\eea
It easy to notice that the $1^-$ and $0^+$ sectors provide degenerate matrices, a result familiar with the corresponding analysis of Einstein theory in second order formulation. Different choices of the non-degenerate $4\times 4$ submatrices are possible and a comforting consistency check reveals that the final shape of the saturated propagator is unchanged. What we think is notable in this kind of gauge invariance checks is that, in some cases, spurious double poles might arise and/or disappear based on the choice of the non-degenerate submatrix. This shows that the physical relevant poles to investigate are the gauge-invariant ones of the saturated propagator, more than those of the inverse of the spin/parity matrices $\tilde b {}^{\left\lbrace S,p \right\rbrace}_{i,j}$. 
To reveal a possible ghost-like propagation we have to follow the protocol illustrated before and arrive at the saturated propagator. In the massless case the check entails asking the propagator to only have single poles and a definite positive residue with the appropriate number of states (see for instance the previous eq.~(\ref{eq15})). Of course in presence of high-rank fields, the large structure of the saturated propagator expanded in components impedes a straightforward check. We will use this example to illustrate the manipulations we adopted in order to reach unambiguous conclusions.\\
Once the gauge constraints (\ref{eq12}) are imposed over the sources the saturated propagator is obtained considering the inverse of \emph{all} the spin/parity sectors $\mathcal D_S(q) = D_S^{3^-}(q) + D_S^{2^+}(q) + D_S^{2^-}(q) + D_S^{1^+}(q) + D_S^{1^-}(q) + D_S^{0^+}(q)$. As said, even the sectors not showing poles have to be included to cancel the spurious ones of the projectors. 
The saturated propagator for the rank-3/rank-2 system has the generic form \cite{Karananas:2014pxa}
\bea \label{eq19}
\lim_{q^2\rightarrow 0}\mathcal D_S = \frac{1}{q^2} 
\begin{pmatrix}
\tilde J^{* \mu \nu} & \tilde q_{\rho}J^{* \mu \nu \rho} 
\end{pmatrix} 
\begin{pmatrix}
\mathcal{M}^{1,1}_{\mu \nu \alpha \beta} &  \mathcal{M}^{1,2}_{\mu \nu \alpha \beta}\\[6pt]
\mathcal{M}^{2,1}_{\mu \nu \alpha \beta} & \mathcal{M}^{2,2}_{\mu \nu \alpha \beta}
\end{pmatrix} 
\begin{pmatrix}
\tilde J^{\alpha \beta} \\[6pt] 
\tilde q_{\iota} J^{\alpha \beta \iota} 
\end{pmatrix} \,.
\eea
In our approach we break Lorentz covariance by expanding (\ref{eq19}) in components, having $q = (\omega, 0, 0, \kappa)$ and imposing $\kappa \rightarrow \omega$ in the massless case. It is therefore more appropriate to collect all the components of $\tilde J^{\mu \nu \alpha}, \tilde J^{\mu \nu}$ (which are less than the unconstrained $J^{\mu \nu \alpha}, J^{\mu \nu}$) in a long row-vector $X = \left(J^{0 1} , J^{0 2}, \cdots \omega J^{0 0 1} , \omega J^{0 0 2} , \cdots\right)$ so that (\ref{eq19}) becomes
\bea \label{eq20}
\lim_{q^2\rightarrow 0}\mathcal D_S = \frac{1}{q^2} \,\,
\tilde X^{\dagger} \mathcal M \tilde X  \,.
\eea
With the matrix $\mathcal M$ we can promptly check the number of propagating states by computing its rank, which, luckily, is equal to two. Then the constraint over the only free parameter $a_{EP}$ is derived diagonalizing $\mathcal M$ which, in our computation, reveals 
\bea \label{eq22}
\lim_{q^2\rightarrow 0}\mathcal D_S = -\frac{1}{a_{EP}}\frac{21 \,|j_1|^2 + 10 \,|j_2|^2}{\omega^2-\kappa^2} \, , 
\eea
with $j_1$ and $j_2$ the two eigenvectors. As expected, the conclusion is that the dimensional $a_{EP}$ coupling allows unitarity if negative. 
\subsection{Singh-Hagen}
The same rationale we introduced in the computation of the mixed rank-3/rank-2 Einstein-Palatini model can help revealing, with some minor changes, the possible presence of ghost and tachyon in models with propagation of massive particles. We will analyze in this and the next section two alternative models which describe a massive spin-3 particle for which the collective cooperation of many different fields seems to be mandatory to preserve locality. The first model was introduced in \cite{Singh:1974qz} and has been recently studied in \cite{Mendonca:2019gco} by integrating out the auxiliary scalar field. We will instead rely on the complete set of operators which we computed for this purpose.\\ 
The Singh-Hagen action is given by 
\bea
S_{SA} &=& \int d^4 x \,\, \bigg[ x_{SA} \bigg(A_{\mu \nu \rho} \Box A^{\mu \nu \rho} - 3 \,\,A_{\mu \nu \rho} \partial^{\mu} \partial_{\alpha} A^{\alpha \nu \rho} - \frac{3}{2} \,A^{\sigma}_{\,\,\nu \sigma} \partial^{\nu} \partial_{\mu} A^{\mu \,\,\rho}_{\,\,\rho} + \nn  \\ 
 && - 3 \, A^{\nu\,\,\sigma}_{\,\,\sigma} \Box A^{\,\,\rho}_{\nu \,\,\rho} + 6\, A_{\rho \sigma}^{\,\,\,\,\,\,\sigma} \partial_{\mu} \partial_{\nu} A^{\mu \nu \rho} + m^2_A \, A^{\mu \nu \rho} A_{\mu \nu \rho} - 3 \, m^2_A\, A^{\mu \rho}_{\,\,\rho} A_{\mu \sigma }^{\,\,\sigma} \bigg) + \nn \\ 
&&+ y_{SA}\,\bigg(\frac{1}{2} \phi \Box \phi + 2\, m_A^2\, \phi^2 \bigg) + z_{SA} \,m_A \, A^{\sigma}_{\,\,\mu \sigma} \partial^{\mu}\phi \bigg]\,\,,
\eea
where $A^{\mu \nu \rho}$ is totally symmetric and we included four free parameters $x_{SA}, y_{SA}, z_{SA}$ and $m_A$ for illustrative reasons. We notice the presence of a term with a quadratic mixing between the rank-3 tensor $A^{\mu \nu \rho}$ and the scalar $\phi$. A symmetric rank-3 tensor contains the following little group representations
\begin{eqnarray}\label{eq23}
&A_{\mu \nu \rho} \supset & 3_1^- \oplus 2_1^+ \oplus 1^-_1 \oplus 1^-_4 \oplus  0^+_1 \oplus 0^+_4 , \end{eqnarray}
which are completed by the trivial one $0^+_8$ carried by the scalar field $\phi$. We then proceed in computing the spin/parity matrices of the Singh-Hagen model which are
\bea \label{eq23a}
a^{\left\lbrace 3,- \right\rbrace}_{i,j} &=& 
-2 \,x_{SA}
\begin{pmatrix}
q^2 - m_A^2 
\end{pmatrix} \, ,  \nn \\ \\
a^{\left\lbrace 2,+ \right\rbrace}_{i,j} &=& 
 2 \,x_{SA} \, m_A^2 \,, \nn \\ \\
a^{\left\lbrace 1,- \right\rbrace}_{i,j} &=& 
x_{SA}
\begin{pmatrix}
8 \left(q^2 - m_A^2\right) & -2 \sqrt{5}\,m_A^2 \\[6pt] 
- 2 \sqrt{5}\,m_A^2 & 0 
\end{pmatrix} \, , \nn \\ 
\eea
and finally  
\bea \label{eq23d}
a^{\left\lbrace 0,+ \right\rbrace}_{i,j} = 
\begin{pmatrix}
 x_{SA} \left(9 q^2 - 4 m_A^2\right) & 3\,x_{SA} \left(q^2 - 2 m_A^2\right) & i\, z_{SA} \,m_A \sqrt{q^2} \\[6pt] 
3\,x_{SA} \left(q^2 - 2 m_A^2\right) & \,x_{SA} \left(q^2 - 4 m_A^2\right) & i\,z_{SA} \,m_A \sqrt{q^2} \\[6pt]
 - i\, z_{SA} \,m_A \sqrt{q^2}  & - i \,z_{SA} m_A \sqrt{q^2}  & -y_{SA} \left(q^2 - 4 m_A^2\right)
\end{pmatrix} \, .
\eea
In the Singh-Hagen model the propagation of the massive spin-3 particle does not require gauge symmetries to prevent the propagation of dangerous ghosts and this is manifested by all the spin/parity matrices being non-degenerate. No constraints are therefore imposed over the sources and we can proceed to the direct inversion of the matrices (\ref{eq23a})-(\ref{eq23d}) to find the saturated propagator  $\mathcal D_S(q) = D_S^{3^-}(q) + D_S^{2^+}(q) + D_S^{1^-}(q) + D_S^{0^+}(q)$. This object inherits the massive pole which is displayed inverting the $a^{\left\lbrace 3,- \right\rbrace}_{i,j}$ matrix. In seeking for the propagating massive states we put $\mathcal D_S(q)$ in a form similar to the massless one (\ref{eq20}). This time we arrange the row-vector as $X = \left(J,  J^{0 0 1} , J^{0 0 2} , \cdots\right)$ and immediately recover the form 
\bea \label{eq24}
\lim_{q^2\rightarrow m_A^2}\mathcal D_S = \frac{1}{q^2-m_A^2} \,\,
X^{\dagger} \mathcal M \, X  \,.
\eea
The matrix $\mathcal M$ has rank 7, which equal the $2\,s + 1$ states of a massive spin-3 particle and the (non zero) eigenvalues are 
\bea
\mathcal M_{res} = x_{SA}^{-1}\,\bigg\{ 3,\,\frac{3}{2},\,\frac{3}{2},\,\frac{3}{2},\,\frac{11}{10},\,\frac{11}{10},\,\frac{11}{10}\bigg\}
\, ,
\eea
which can be made positive if we keep $x_{SA} > 0$. As already discussed in \cite{Mendonca:2019gco} not all the parameters we used are physical and only $x_{SA}$ and $m_A$ are constrained by causality and unitarity. 
\section{Applications: the spectrum of the Klishevich-Zinoviev model} \label{KZ}
The Singh-Hagen model is one of the primitive attempts to define a linear Lagrangian for higher-spin particles. Further efforts to build a full dynamic around these germinal models were met with the numerous constraints which upset the definition of a non-trivial S-matrix (see, for instance, \cite{Porrati:2008rm,Bekaert:2010hw} and references therein). In order to overcome the same constraints, alternative approaches have been investigated, a notable one being Vasyliev's full non-linear theory \cite{Vasiliev:1990en,Vasiliev:1995dn,Vasiliev:2000rn}.
The presence, in these models, of a non-zero cosmological constant introduces a series of peculiar phenomena like novel stability constraints and the appearance of partial masslessness for propagating particles \cite{Deser:2001us,Deser:2001xr,Deser:2001pe,Deser:2001wx,Breitenlohner:1982jf}. 
In particular, in order to investigate partially massless theories in Anti de Sitter spaces, a convenient Lagrangian setup describing massive higher-spin propagation has been developed in \cite{Klishevich:1997pd,Zinoviev:2008ck,Zinoviev:2008ze}. For our purposes, it is notable that such formulation admits a flat limit and contributes for an alternative description of higher-spin particles in Minkowski space. Moreover, differently from the Singh-Hagen case, this new formulation is inherently gauge invariant, requiring all the set of auxiliary Stueckelberg fields with rank $\leq 3$ to mix in the quadratic action. This model is therefore the perfect arena to test our operators and provide an alternative check of unitarity and causality violations where, moreover, shortcut like (\ref{eq14b}) would be ineffective.  \\ 
Again, we rely on totally symmetric tensors so that the set of little group representations carried by the fields are
\begin{eqnarray}\label{eq25}
A_{\mu \nu \rho} &\supset & 3_1^- \oplus 2_1^+ \oplus 1^-_1 \oplus 1^-_4 \oplus  0^+_1 \oplus 0^+_4 , \nn \\
H_{\mu \nu} &\supset & 2_4^+ \oplus 1^-_7 \oplus 0^+_5 \oplus 0^+_6 , \nn \\
V_{\mu} &\supset &  1^-_8 \oplus 0^+_7 \, , \nn \\
\phi &\supset & 0^+_8 \,.
\end{eqnarray} 
In terms of these fields the action can be written as a sum of distinctive parts 
\bea
S_{KZ} = S_{AA} + S_{HH} + S_{VV} + S_{\phi \phi} + S_{mix} \, ,
\eea
with
\bea
S_{AA} &=& \int d^4 x \bigg[ \bigg( -\frac{1}{2}\partial_{\alpha}A_{\mu \nu \rho}\,\partial^{\alpha}A^{\mu \nu \rho} + \frac{3}{2} \partial^{\mu}A_{\mu \nu \rho} \partial^{\alpha}A_{\alpha}^{\,\,\nu \rho} + \frac{3}{2} \partial^{\alpha} A_{\mu}^{\,\,\mu \nu} \partial_{\alpha}A_{\rho\,\,\nu}^{\,\,\rho} + \nn \\ &&+ \frac{3}{4}\partial_{\nu}A_{\mu}^{\,\,\mu \nu} \partial_{\alpha}A_{\rho}^{\,\,\rho \alpha} - 3 \partial_{\nu}A_{\sigma\,\,\rho}^{\,\,\sigma} \partial_{\mu} A^{\mu \nu \rho} + \bigg) + m_A^2 \bigg(\frac{1}{2} A^{\mu \nu \rho} A_{\mu \nu \rho} - \frac{3}{2} A^{\mu \, \, \alpha}_{\,\,\alpha} A_{\mu \beta}^{\,\,\,\,\,\,\beta} \bigg) \bigg] \, , \nn  \\ \\
S_{HH} &=& \int d^4 x \bigg[ 3 \bigg(\frac{1}{2}\partial^{\mu} H^{\alpha \beta} \partial_{\mu} H_{\alpha \beta} - \partial_{\beta} H^{\mu \beta} \partial_{\alpha}H_{\mu}^{\,\,\alpha} + \partial_{\beta}H^{\mu \beta}\partial_{\mu} H_{\alpha}^{\,\,\alpha} - \frac{1}{2} \partial_{\mu} H^{\alpha}_{\,\,\alpha}\partial^{\mu} H^{\beta}_{\,\,\beta} \bigg) + \nn \\ &+&  \frac{9\,m_A^2}{4}  H^{\alpha}_{\,\,\alpha} H^{\beta}_{\,\,\beta}  \bigg] \, , \\
S_{VV} &=&  \int d^4 x  \bigg[ - \frac{15}{2} \bigg(\partial^{\mu} V^{\nu} \partial_{\mu} V_{\nu} - \partial^{\mu} V^{\nu} \partial_{\nu} V_{\mu} \bigg) - \frac{45\,m_A^2}{4} V_{\mu} V^{\mu} \bigg] \, , \\
S_{\phi \phi} &=& \int d^4 x  \bigg(-\frac{45}{2} \partial^{\mu} \phi \, \partial_{\mu} \phi + 225 \,m_A^2\, \phi^2
\bigg)  \, ,
\eea
and 
\bea
S_{mix} &=& \int d^4 x \bigg[ m_A\,\bigg( -\frac{3}{2} \big( 2 \, A^{\mu \nu \alpha}\partial_{\mu}H_{\nu \alpha} - 4 \, A^{\alpha \,\,\mu}_{\,\,\alpha} \partial_{\beta}H^{\beta}_{\,\,\mu} + A^{\alpha \,\, \mu}_{\,\,\alpha} \partial_{\mu} H_{\beta}^{\,\,\beta} \big) + 15 \, \big(H^{\mu \nu}\partial_{\mu}V_{\nu} - H^{\alpha}_{\,\,\alpha} \partial_{\mu}V^{\mu} \big) + \nn \\  &-& 225 \, V^{\mu}\partial_{\mu}\phi \bigg) + m_A^2\,\bigg( \frac{15}{2}\, A^{\mu\,\,\alpha}_{\,\,\alpha}V_{\mu} - 45 H^{\alpha}_{\,\,\alpha} \phi  \bigg) \bigg] \, .
\eea
Normalizations of the various terms are taken from \cite{Zinoviev:2008ck}.\\ The four spin/parity matrices are
\bea \label{eq25a}
&&a^{\left\lbrace 3,- \right\rbrace}_{i,j} = m_A^2 - q^2 \, , \\ \nn \\
&&a^{\left\lbrace 2,+ \right\rbrace}_{i,j} = \begin{pmatrix}
 m_A^2 & - i \sqrt{3} m_A \sqrt{q^2} \\[6pt]
 i \sqrt{3} m_A \sqrt{q^2} & 3 q^2 
\end{pmatrix}  \,, \\ \nn \\
&&a^{\left\lbrace 1,- \right\rbrace}_{i,j} = \begin{pmatrix}
 4\left( q^2 - m_A^2\right) & -\sqrt{5} \, m_A^2 & i\, \sqrt{30}\, m_A \, \sqrt{q^2} & + m_A^2 \, \frac{5\,\sqrt{15}}{2}  \\[6pt]
 -\sqrt{5}\, m_A^2 & 0 & 0 &  m_A^2 \, \frac{5\,\sqrt{3}}{2}  \\[6pt] 
- i \, \sqrt{30} \, \sqrt{q^2} & 0 & 0 & i \, m_A^2 \, \frac{15}{\sqrt 2}  \\[6pt]
m_A^2\,\frac{5\,\sqrt{15}}{2} & m_A^2\,\frac{5\,\sqrt{3}}{2} & - i \, m_A^2 \frac{15}{\sqrt 2} & - \frac{15}{2}\left(2 q^2 + 3 m_A^2\right)   
\end{pmatrix} \,, \\ \nn \\ 
&&a^{\left\lbrace 0,+ \right\rbrace}_{i,j} = \nn \\
&&\begin{pmatrix}
 \frac{9}{2} \, q^2 - 2 \, m_A^2  & \frac{3}{2}\, q^2 - 3 \, m_A^2 &  - m_A\, i\, \frac{5\,\sqrt{3}}{2} \sqrt{q^2} & + m_A \,\frac{9\,\,i}{2} \sqrt{q^2} & \frac{15}{2} \, m_A & 0 \\[6pt]
\frac{3}{2} \, q^2 - 2 \, m_A^2 & \frac{1}{2} \, q^2 - 2 \, m_A^2 &  - m_A\, i\, \frac{3\,\sqrt{3}}{2} \sqrt{q^2} & + m_A \frac{3\,\,i}{2} \sqrt{q^2} & \frac{15}{2} m_A & 0 \\[6pt]
m_A \,i\,\frac{5 \sqrt{3}}{2}\sqrt{q^2} & m_A \,i\, \frac{3\sqrt{3}}{2} \sqrt{q^2} & \frac{27}{2}\, m_A^2 - 6 \, q^2 & \frac{9 \sqrt{3}}{2}\, m_A^2 & - i \, 15 \, \sqrt{3} \, m_A \, \sqrt{q^2} & - 45 \, \sqrt{3} \, m_A^2  \\[6pt]
m_A \,i\,\frac{9}{2}\sqrt{q^2} & - m_A \,i\, \frac{3}{2} \, \sqrt{q^2} & \frac{9 \sqrt{3}}{2} \, m_A^2 & \frac{9}{2} \, m_A^2 & 0 & - 45 \,  m_A^2  \\[6pt]
m_A^2 \, \frac{15}{2} & m_A^2 \, \frac{15}{2} & 15  \,\sqrt{3} \, i \, m_A \, \sqrt{qq^2} & 0 & -\frac{45}{2}\,m_A^2 & - 225\, i\, m_A \,\sqrt{q^2}  \\[6pt]
0 & 0 & -45 \, \sqrt{3} \,  m_A^2  & - 45 \, m_A^2  & 225 \,i \, m_A \, \sqrt{q^2} & 90 \, \left(q^2 + 5 m_A^2\right) 
\end{pmatrix}  \nn \\\nn \\
\eea
The pervasiveness of gauge invariance is revealed noticing that the rank of the matrices $a^{\left\lbrace 2,+ \right\rbrace}_{i,j}, a^{\left\lbrace 1,- \right\rbrace}_{i,j}$ and $a^{\left\lbrace 0,+ \right\rbrace}_{i,j}$ is, respectively, $1,2,5$. The set of null-vectors will therefore generate a long list of constraints which, again, we will solve for the source components in the frame $q = (\omega, \vec{0} )$.
After this step we can again discover the structure eq.(\ref{eq24}) with the eigenvalues of the large matrix $M$ being 
\bea
\mathcal M_{res} &=& \bigg\{ \frac{1}{15}\left(533 - 2 \sqrt{61591}\right),\, 
\frac{1}{15}\left(533 - 2 \sqrt{61591}\right), \,
\frac{1}{15}\left(533 + 2 \sqrt{61591}\right),\, \frac{1}{15}\left(533 + 2 \sqrt{61591}\right), \nn \\ 
&& 3,\,5,\,\frac{683}{45}\bigg\}\nn
\, ,
\eea
which, while hideous to look at, are all positive. 
\section{Applications: collective propagation for higher-spin particles}
The previous examples displayed how the quadratic mixing between fields of different rank helps modeling manifestly local higher-spin propagation. It is therefore natural to expect that the corresponding Lagrangians might lead to a broader spectrum than that of a single particle, and indeed this is the direction taken by the numerous studies around extensions of the minimal Einstein-Palatini model of section (\ref{secEP}), see for instance \cite{Percacci:2020ddy,Neville:1978bk,Neville:1979rb,Sezgin:1979zf,Karananas:2014pxa,Sezgin:1981xs,Lin:2018awc,Lin:2020phk,Alvarez:2018lrg,Aoki:2019rvi,Blagojevic:2018dpz,Baikov:1992uh}. \\
The full set of operators, completed by this work, opens a computational opportunity for the derivation of parameter constraints for such collective systems. It is outside of this paper's purpose to provide a survey of new models achievable with the projectors formalism, an activity whose extent asks for dedicated future efforts. 
Nevertheless, for the sake of completeness, we conclude by showing an explicit, and seemingly new, model with collective higher-spin propagation of massive states. \\
It has to be clarified, once again, that the shaping of a unitary and causal Lagrangian is a first step which needs to be supported by the modeling of possible consistent interactions. This latter process might relegate the linear theory to irrelevance in presence of obstructions, confining the theory to the dull status of a free one. The building of consistent interactions on top of the free theory calls for a meticulous analysis which, again, evades the scope of our work. 
\subsection{Collective modes in massive spin-3/spin-1 propagation}
With the priviledge of arbitrariety we direct our focus to the spectrum of the mixed system for the three fields 
\begin{eqnarray}\label{eq26}
A_{\mu \nu \rho} &\supset & 3_1^- \oplus 2_1^+ \oplus 1^-_1 \oplus 1^-_4 \oplus  0^+_1 \oplus 0^+_4 , \nn \\
V_{\mu} &\supset &  1^-_8 \oplus 0^+_7 \, , \nn \\
\phi &\supset & 0^+_8 \,,
\end{eqnarray} 
setting our goal in finding a propagating spin-3 particle and, possibly, other accompanying healty massive particles. By starting with all the parameters unconstrained we are forced to repeat, for each of our attempts, the process illustrated in the previous sections leading, ultimately, to the saturated propagator. In every case we would repeat such steps and, in presence of pathologies, we would seek for a coupling texture in order to prevent them. To avoid redundancies and the corresponding tenfold growth of the volume of this paper we will refrain, when not strictly necessary, to show all the details of the process.\\
As in the previous examples we want to study the action
\bea
S= S_{AA} +  S_{VV} + S_{\phi \phi} + S_{mix} \, ,
\eea
which, at the beginning, has the generic unconstrained form defined by
\bea
S_{AA} &=& \int d^4 x \bigg[  a_1\,\partial_{\alpha} A_{\mu \,\, \sigma}^{\,\, \sigma} \partial^{\mu}A^{\rho \,\, \alpha}_{\,\,\rho} + a_2 \, \partial_{\alpha} A_{\nu \,\, \sigma}^{\,\,\sigma} \partial^{\alpha} A^{\mu \,\, \nu}_{\,\,\mu } + a_3\, \partial_{\alpha} A^{\alpha \mu \nu} \partial_{\beta}A_{\mu \nu}^{\,\,\,\,\beta} + \nn \\ && + a_4 \, \partial^{\nu} A^{\alpha \,\, \beta}_{\,\,\alpha} \partial_{\beta}A_{\beta \nu}^{\,\,\,\, \mu} + a_7 \, \partial_{\nu}A_{\mu \nu \rho} \partial^{\nu} A^{\mu \nu \rho}  + m_1^2 \, A^{\mu \nu \rho} A_{\mu \nu \rho} + m_2^2 \, A^{\mu \,\, \nu }_{\,\,\mu} A_{\nu \,\, \rho}^{\,\, \rho}  \bigg] \, , \nn \\
S_{VV} &=&  \int d^4 x  \bigg[ v_1 \partial^{\mu} V^{\nu} \partial_{\mu} V_{\nu} + v_2 \,\partial^{\mu} V^{\nu} \partial_{\nu} V_{\mu}  + m^2_V V_{\mu} V^{\mu} \bigg] \, , \nn \\
S_{\phi \phi} &=& \int d^4 x  \bigg( s_1 \partial^{\mu} \phi \, \partial_{\mu} \phi + \,m_{S}^2\, \phi^2
\bigg)  \, ,
\eea
and 
\bea
S_{mix} &=& \int d^4 x \bigg[ m_{AS} \, \phi \, \partial_{\beta} A^{\alpha \,\, \beta}_{\,\,\alpha} + m_{AV} \, V_{\mu} \, A^{\mu \alpha}_{\,\,\,\,\,\,\alpha} + m_{VS}\, \phi \, \partial_{\alpha} V^{\alpha} \bigg] \, .
\eea
The spin/parity matrices are easily found to be
\bea \label{eq26a}
&&a^{\left\lbrace 3,- \right\rbrace}_{i,j} =  2 \left( m_1^2 + a_7  \, q^2 \right)\, , \nn \\ \\
&&a^{\left\lbrace 2,+ \right\rbrace}_{i,j} =  2 \left[ m_1^2 + \left(a_7 + \frac{a_3}{3} \right) \, q^2 \right]\, ,\nn \\ \\
&&a^{\left\lbrace 1,- \right\rbrace}_{i,j} = \nn \\ 
&&\begin{pmatrix}
 \frac{2}{3}\left(3 m_1^2 + 5 m_2^2 + q^2 \left(5 a_2 + 3 a_7\right) \right) &  \frac{\sqrt{5}}{3}\left(2 m_2^2 + q^2 \left(2 a_2 + a_4\right) \right) & \sqrt{\frac{5}{3}} m_{AV}\, \\[6pt]
 \frac{\sqrt{5}}{3}\left(2 m_2^2 + q^2 \left(2 a_2 + a_4\right) \right) & \frac{2}{3}\left(3 m_1^2 +  m_2^2 + q^2 \left(a_2 + 2 a_3 + a_4 + 3 a_7\right) \right) &   \frac{m_{AV}}{\sqrt{3}}\\[6pt] 
\sqrt{\frac{5}{3}}\,m_{AV} & \frac{m_{AV}}{\sqrt{3}} & 2 \left(m_V^2 + v_1 q^2 \right)   
\end{pmatrix} \,, \nn \\
\eea
and 
\bea \label{eq26d}
&&\hspace{-3.5cm} a^{\left\lbrace 0,+ \right\rbrace}_{i,j} = \nn \\
&&\hspace{-3.6cm}\begin{pmatrix}
 2\,\left(m_1^2 + m_2^2 + q^2 \left(a_1 + a_2 + \frac{a_3}{3} + a_7 \right) \right) &  2\,\left( m_2^2 + q^2 \left(a_1 + a_2 + \frac{a_4}{2} \right) \right)  & m_{AV} & - i m_{AS} \sqrt{q^2}\, \\[6pt]
 2\,\left( m_2^2 + q^2 \left(a_1 + a_2 + \frac{a_4}{2} \right) \right) &  2\,\left(m_1^2 + m_2^2 + q^2 \left(a_1 + a_2 + a_3 + a_4 + a_7 \right) \right)  & m_{AV} & - i m_{AS} \sqrt{q^2}\, \\[6pt]
 m_{AV} & m_{AV}  & 2 \left( m_V^2 + q^2 \left(v_1 + v_2\right) \right) & - i m_{AS} \sqrt{q^2}\, \\[6pt]
i m_{AS} \sqrt{q^2} & i m_{AS} \sqrt{q^2}  & i m_{AS} \sqrt{q^2} & 2 \left(m_S^2 + s_1 \, q^2 \right)\,  
 \end{pmatrix} \,. \nn \hspace{-1cm} \\ \nn \\ 
\eea 
We notice immediately that a unique mass parameter appears in the combinations defining the masses of the $3^-$ and $2^+$ sector. Once we compute the saturated propagator above the two massive poles $m^2_{3^-} = -\frac{a_7}{m_1^2}$ and $m^2_{2^+} = -\frac{a_7+a_3/3}{m_1^2}$ we find, as common when spin sectors stem from the same field, that is impossible to simultaneously require causality and unitarity for both (see also \cite{Anselmi:2020opi}). We solve this by requiring 
\bea
a_7 = -1\,, \,\, m^2_1 > 0 \,\,\text{and}\,\, a_3 = 3 \, .
\eea
These conditions will allow the non pathological propagation of a massive spin-3 particle and make the $2^+$ sector innocuous. For the larger $1^-$ and $0^+$ sectors the first obstacle is represented by higher order poles $q^{2 n}$ which show up already in computing the determinant of the corresponding spin/parity matrices. In looking for physical propagation we can explore different ways to reduce the degree of such polynomials in $q^2$ and check, for each possibility, if they welcome unitarity and causality. After trials and errors we found a workable parameter space by asking the poles of the $0^+$ sector to disappear and $1^-$ to have a single massive pole in $q^2$. These requirements define a limited set of different parameter textures which we can explore in full detail. Again, this choice was dictated by an apparent simplicity of the constraints and is in no way unique. In principle is possible to let the determinant being higher order in $q^2$ without this translating  in higher order poles of the propagator (for instance, $n$ massive scalar fields would generate a $q^{2n}$ polynomial in the $a^{\left\lbrace 0,+ \right\rbrace}_{i,j}$ determinant). Moreover, to coherently follow the initial choice of model's degrees of freedom  $A_{\mu \nu \rho}, V_{\mu}$ and $\phi$, we have only explored cases with  non-zero quadratic mixing. \\
Once we narrowed the parameter space, the constraining power of unitarity and causality can fully unfold and remove away all but one surviving possibility.   
This is greatly simplified by setting the free parameter $a_4 = 0$ and getting, by simply solving the algebrical equations that nullify the unwanted coefficients of higher order poles, the restrictions: 
\bea
a_1 = -\frac{3}{4}, \,\, a_2 = \frac{3}{4},\,\, m^2_2 = - \frac{9}{8} m_1^2, \,\, m_{AS} = \frac{m^2_S - 4 m_1^2s_1}{4 \sqrt{2 m_S^2}}, \,\, m_V^2 = 0, \,\, m_{AV} = -\frac{m_1^2 m_{VS}}{\sqrt{2 m_S^2}} 
\eea
Once these conditions are adopted the saturated propagator over the mass   
\bea
m^2_{1^-} = -\frac{3 \, m_1^2 \,m^2_{VS}}{2\, m_{VS}^2 + 15\, m_S^2\, v_1}  
\eea
can be promtly computed. We find three propagating states with same residue $Res_{q^2 \rightarrow m^2_{1^-}}$
\bea
Res_{q^2 \rightarrow  m^2_{1^-}} =  - \frac{3\,\left(56\, m_{VS}^6 + 5625\, m_S^6 \,v_1^2 + 20 \,m_{VS}^4\, m_S^2 \left( 5 + 6 v_1\right) + 300 \,m_{VS}^2 \,m_S^4 \, v_1 \, \left( 5+ 6 v_1 \right) \right)}{10 \left( 2\, m_{VS}^2 + 15\, m _S^2 \,v_1 \right)^3} \, .\nn \\
\eea
Solutions which are causal and unitary exist provided that 
\bea
m_{VS} \neq 0 , \,\, m_V^2 < 0, \,\, m_S^2 > 0, \,\, v_1 < 0, m_1^2 >0, \, \, \text{and}\,\, 2 \, m_{VS}^2 + 15\, m_S^2\, v_1 < 0 \, . 
\eea

\section{Conclusions}
With this work we have completed the full set of operators needed to uncover the propagating spin/parity components expressed in quadratic Lagrangians of tensor fields up to rank-3. We did so by supporting the already known projectors with the tensor functions connecting different representations of the same spin/parity sector.
We have illustrated how a dedicated sequence of manipulations can help to assess, unambiguously and in a gauge invariant way, the physical properties of the particle quanta. By focusing on the saturated propagator all possible cases, massless as well as critical ones, can be tackled and the only drawback is represented by the inversion of large matrices, when multiple components belong to the same sector. 
With such machinery we have then studied the spectrum of well known theories exhibiting quadratic mixing, like Stueckelberg and Einstein gravity in Palatini formalism, as well as more exotic ones like the Singh-Hagen model and the gauge-based Klishevich-Zinoviev system.
The latter for the first time subjected to a spectral analysis based on the use of projectors algebra. \\
Finally, after a long scrutiny through the strong constraints of ghost and tachyon elusion, we have defined a new model collectively propagating a massive vector together with a massive spin-3 particle. 
%

\subsection*{Acknowledgments}
We thank Roberto Percacci and Marco Piva for helpful clarifications. \\
This work was supported by the Estonian Research Council grant PRG803
and by the EU through the European Regional Development Fund
CoE program TK133 ``The Dark Side of the Universe."

\begin{appendix}

\section{Spin-projector operators including rank-0 and rank-1 fields}
We list here the independent set of spin-projector operators which completes the ones of \cite{Mendonca:2019gco,Percacci:2020ddy} by including representations carried by scalar and vector fields. The operators missing from this list can be recovered via eq.~(\ref{eq6}). 
\subsection{$\bm{ P^{0,p}_{\lbrace i=1\cdots7,7 \rbrace}}$}
\begin{align}
& P^{0,p}_{\lbrace 1,7 \rbrace}{}^{\alpha \beta \gamma}{}_{\mu} = \frac{\bigl(q^2 g^{\beta \gamma} q^{\alpha} + q^2 g^{\alpha \gamma} q^{\beta} + (q^2 g^{\alpha \beta} - 3 \, q^{\alpha} q^{\beta}) q^{\gamma}\bigr) q_{\mu}}{3\, q^4},& \hspace{3cm} \nn \\
&  P^{0,p}_{\lbrace 2,7\rbrace}{}^{\alpha \beta \gamma}{}_{\mu} = \frac{(2\, g^{\beta \gamma} q^{\alpha} -  g^{\alpha \gamma} q^{\beta} - g^{\alpha \beta} q^{\gamma}) q_{\mu}}{3\, \sqrt{2} q^2},& \hspace{3cm} \nn \\
& P^{0,p}_{\lbrace 3,7\rbrace}{}^{\alpha \beta \gamma}{}_{\mu} = \frac{(g^{\alpha \gamma} q^{\beta} -  g^{\alpha \beta} q^{\gamma}) q_{\mu}}{\sqrt{6}\, q^2},& \hspace{3cm} \nn \\ 
& P^{0,p}_{\lbrace 4,7\rbrace}{}^{\alpha \beta \gamma}{}_{\mu} = \frac{q^{\alpha} q^{\beta} q^{\gamma} q_{\mu}}{q^4},& \hspace{3cm} \nn \\
&  P^{0,p}_{\lbrace 5,7\rbrace}{}^{\alpha \beta}{}_{\mu} = \frac{(q^2 g^{\alpha \beta} - q^{\alpha} q^{\beta}) q_{\mu}}{\sqrt{3}\, (q^2)^{3/2}}, & \hspace{3cm} \nn \\ 
& P^{0,p}_{\lbrace 6,7\rbrace}{}^{\alpha \beta}{}_{\mu} = \frac{q^{\alpha} q^{\beta} q_{\mu}}{(q^2)^{3/2}}, & \hspace{3cm} \nn \\ 
&  P^{0,p}_{\lbrace 7,7\rbrace}{}^{\alpha}{}_{\beta} = \frac{q^{\alpha} q_{\beta}}{q^2}. & \hspace{3cm} 
\end{align}
\subsection{$\bm{ P^{1,m}_{\lbrace i=1\cdots 8,8 \rbrace}}$}
\bea
&& P^{1,m}_{\lbrace 1,8\rbrace}{}^{\alpha \beta \gamma}{}_{\mu} = \frac{1}{\sqrt{15} q^4} \bigg(q^4 g^{\alpha \beta} \delta^{\gamma}_{\mu} -  q^2 \delta^{\gamma}_{\mu} q^{\alpha} q^{\beta} -  q^2 \delta^{\beta}_{\mu} q^{\alpha} q^{\gamma} + q^2 \delta^{\alpha}_{\mu} (q^2 g^{\beta \gamma} -  q^{\beta} q^{\gamma}) -  q^2 g^{\beta \gamma} q^{\alpha} q_{\mu} -  q^2 g^{\alpha \beta} q^{\gamma} q_{\mu} + \nn \\ && +  3\, q^{\alpha} q^{\beta} q^{\gamma} q_{\mu} + q^2 g^{\alpha \gamma} (q^2 g^{\beta}_{\mu} -  q^{\beta} q_{\mu})\bigg), \nn \\
&& P^{1,m}_{\lbrace 2,8\rbrace}{}^{\alpha \beta \gamma}{}_{\mu} = \frac{1}{2 \sqrt{3} \,q^2} \bigg(q^2 g^{\alpha \beta} \delta^{\gamma}_{\mu} -  \delta^{\gamma}_{\mu} q^{\alpha} q^{\beta} -  \delta^{\beta}_{\mu} q^{\alpha} q^{\gamma} - 2\, \delta^{\alpha}_{\mu} (q^2 g^{\beta \gamma} -  q^{\beta} q^{\gamma}) + \nn \\ && + 2\, g^{\beta \gamma} q^{\alpha} q_{\mu} -  g^{\alpha \beta} q^{\gamma} q_{\mu} -  g^{\alpha \gamma} (- q^2 \delta^{\beta}_{\mu} + q^{\beta} q_{\mu})\bigg), \nn \\
&& P^{1,m}_{\lbrace 3,8\rbrace}{}^{\alpha \beta \gamma}{}_{\mu} = \frac{q^{\alpha} (- \delta^{\gamma}_{\mu} q^{\beta} + \delta^{\beta}_{\mu} q^{\gamma}) + g^{\alpha \gamma} (- q^2 g^{\beta}_{\mu} + q^{\beta} q^{\mu}) + g^{\alpha \beta} (q^2 \delta^{\gamma}_{\mu} -  q^{\gamma} q_{\mu})}{2 q^2}, \nn \\
&& P^{1,m}_{\lbrace 4,8\rbrace}{}^{\alpha \beta \gamma}{}_{\mu} = \frac{q^2 \delta^{\gamma}_{\mu} q^{\alpha} q^{\beta} + q^{\gamma} \bigl(q^2 \delta^{\beta}_{\mu} q^{\alpha} + q^{\beta} (q^2 \delta^{\alpha}_{\mu} - 3\, q^{\alpha} q_{\mu})\bigr)}{\sqrt{3}\, q^4},\nn \\ 
&& P^{1,m}_{\lbrace 5,8\rbrace}{}^{\alpha \beta \gamma}{}_{\mu} = \frac{\delta^{\gamma}_{\mu} q^{\alpha} q^{\beta} + \delta^{\beta}_{\mu} q^{\alpha} q^{\gamma} - 2\, \delta^{\alpha}_{\mu} q^{\beta} q^{\gamma}}{\sqrt{6} q^2}, \nn \\
&& P^{1,m}_{\lbrace 6,8\rbrace}{}^{\alpha \beta \gamma}{}_{\mu}=\frac{q^{\alpha} (\delta^{\gamma}_{\mu} q^{\beta} - \delta^{\beta}_{\mu} q^{\gamma})}{\sqrt{2}\, q^2},\nn \\ 
&& P^{1,m}_{\lbrace 7,8\rbrace}{}^{\alpha \beta}{}_{\mu} = \frac{q^2 \delta^{\beta}_{\mu} q^{\alpha} + q^{\beta} (q^2 \delta^{\alpha}_{\mu} - 2\, q^{\alpha} q_{\mu})}{\sqrt{2}\, (q^2)^{3/2}}, \nn \\
&& P^{1,m}_{\lbrace 8,8\rbrace}{}^{\alpha}{}_{\beta} = \delta^{\alpha}_{\beta} -  \frac{q^{\alpha} q_{\beta}}{q^2} .
\eea
\subsection{$\bm{ P^{0,p}_{\lbrace i=1\cdots 8,8 \rbrace}}$}
\begin{align}
& P^{0,p}_{\lbrace 1,8\rbrace}{}^{\alpha \beta \gamma}=\frac{q^2 g^{\beta \gamma} q^{\alpha} + q^2 g^{\alpha \gamma} q^{\beta} + (q^2 g^{\alpha \beta} - 3 \,q^{\alpha} q^{\beta}) q^{\gamma}}{3\, (q^2)^{3/2}}, & \hspace{3cm} \nn \\ 
& P^{0,p}_{\lbrace 2,8\rbrace}{}^{\alpha \beta \gamma} = \frac{2 \,g^{\beta \gamma} q^{\alpha} -  g^{\alpha \gamma} q^{\beta} - g^{\alpha \beta} q^{\gamma}}{3\, \sqrt{2} (q^2)^{1/2}}, & \hspace{3cm} \nn \\ 
& P^{0,p}_{\lbrace 3,8\rbrace}{}^{\alpha \beta \gamma} = \frac{g^{\alpha \gamma} q^{\beta} -  g^{\alpha \beta} q^{\gamma}}{\sqrt{6}\, (q^2)^{1/2}}, \nn \\ 
& P^{0,p}_{\lbrace 4,8\rbrace}{}^{\alpha \beta \gamma} = \frac{q^{\alpha} q^{\beta} q^{\gamma}}{(q^2)^{3/2}}, & \hspace{3cm} \nn \\
& P^{0,p}_{\lbrace 5,8\rbrace}{}^{\alpha \beta} = \frac{q^2 g^{\alpha \beta} - q^{\alpha} q^{\beta}}{\sqrt{3} q^2}, & \hspace{3cm} \nn \\
& P^{0,p}_{\lbrace 6,8\rbrace}{}^{\alpha \beta} = \frac{q^{\alpha} q^{\beta}}{q^2}, & \hspace{3cm}\nn \\
& P^{0,p}_{\lbrace 7,8\rbrace}{}^{\alpha} = \frac{q^{\alpha}}{(q^2)^{1/2}}, & \hspace{3cm} \nn \\
& P^{0,p}_{\lbrace 8,8\rbrace} = 1 . & \hspace{3cm}
\end{align}

\end{appendix}


\end{document}